\documentclass{kapproc} 
\usepackage{procps}
\usepackage[dvips]{graphicx}
\upperandlowercase
\kluwerbib 
\begin{document}
%
%
\articletitle[Bar-Driven Evolution]
{Bar-Driven Evolution and \\2D Spectroscopy of Bulges}
\author{M.\ Bureau\altaffilmark{1}, E.\ Athanassoula\altaffilmark{2},
A.\ Chung\altaffilmark{3} and G.\ Aronica\altaffilmark{4}}
\affil{\altaffilmark{1}Hubble Fellow, Columbia Astrophysics
 Laboratory, 550 West 120th Street, 1027 Pupin Hall, MC~5247, New
 York, NY~10027, U.S.A.,\\ \altaffilmark{2}Observatoire de Marseille,
 2 place Le Verrier, F-13248 Marseille Cedex~4, France,\\
 \altaffilmark{3}Department of Astronomy, Columbia University, 550
 West 120th Street, 1411 Pupin Hall, MC~5246, New York, NY~10027,
 U.S.A.,\\ \altaffilmark{4}Astronomisches Institut,
 Ruhr-Universit$\ddot{\textrm{\small\it a}}$t Bochum, D-44780 Bochum, Germany}
%
%
\begin{abstract}
A multi-faceted approach is described to constrain the importance of
bar-driven evolution in disk galaxies, with a special emphasis on
bulge formation. N-body simulations of bars are compared to the
stellar kinematics and near-infrared morphology of $30$ edge-on
spirals, most with a boxy bulge. The N-body simulations allow to
construct stellar kinematic bar diagnostics for edge-on systems and to
quantify the expected vertical structure of bars. Long-slit spectra of
the sample galaxies show characteristic double-hump rotation curves,
dispersion profiles with secondary peaks and/or flat maxima, and
correlated $h_3$ and $V$ profiles, indicating that most of them indeed
harbor edge-on bars. The stellar kinematics also suggests the presence
of cold, quasi-axisymmetric central stellar disks. The ionized-gas
distribution and kinematics further suggests that those disks formed
through bar-driven gaseous inflow and subsequent star formation, which
are absent in our simulations. Minimally affected by dust and
dominated by Population~II stars, $K$-band imaging of the same
galaxies spectacularly highlights radial variations of the bars'
scaleheights, as expected from vertical disk instabilities. The light
profiles also vary radially in shape but never approach a classic
deVaucouleurs law. Filtering of the images further isolates the
specific orbit families at the origin of the boxy structure, which can
be directly related to periodic orbit calculations in generic 3D
barred potentials. Bars are thus shown to contribute substantially to
the formation of both large-scale triaxial bulges and embedded central
disks. Relevant results from the {\tt SAURON} survey of the
stellar/ionized-gas kinematics and stellar populations of spheroids
are also briefly described. Specific examples supporting the above
view are used to illustrate the potential of coupling stellar
kinematics and linestrengths (age and metallicity), here specifically
to unravel the dynamical evolution and related chemical enrichment
history of bars and bulges.
\end{abstract}
%
%
\section[Introduction]{Introduction\label{sec:intro}}
Conventional wisdom states that the bulges of spiral galaxies are
analogous to low-luminosity elliptical galaxies residing at the
centers of disks (Davies et al.\ 1983), and thus probably formed
through rapid collapse or merging. There is however mounting evidence
against significant merger growth in many bulges and numerous studies
argue for the importance of slower (i.e.\ secular) processes (e.g.\
Kormendy 1993; Andredakis, Peletier, \& Balcells 1995). Given the
rapid decrease of the merger and star formation rates since
$z\approx1$, secular evolution mechanisms have probably been
non-negligible for some time already, and their relative importance
will only increase with time. Most works emphasize the potentially
crucial role of bars and other asymmetries in disks (e.g.\ Friedli et
al.\ 1996; Erwin \& Sparke 2002). Theoretical models often involve the
growth of a central mass through bar-driven inflow and recurring bar
destruction and formation (e.g.\ Pfenniger \& Norman 1990; Friedli \&
Benz 1993). The efficiency of bar dissolution mechanisms remains
however uncertain and bar re-formation requires substantial external
gas accretion over the lifetime of a galaxy (e.g.\ Bournaud \& Combes
2002; Shen \& Sellwood 2004). In this paper, we thus focus on slow
processes which can occur in isolation.

Beside bar formation, which is well-studied and documented, N-body
simulations of cold disks systematically show that, soon after their
formation, bars should thicken and settle with an increased vertical
velocity dispersion, appearing boxy or peanut-shaped (B/PS) when
viewed edge-on (e.g.\ Combes \& Sanders 1981; Combes et al.\
1990). The large but constant fraction of B/PS bulges across the
Hubble sequence ($\approx45\%$) supports both the importance of this
mechanism for real galaxies and its association with bars (e.g.\
L$\ddot{\textrm{u}}$tticke, Dettmar, \& Pohlen 2000). Although bars
are not readily identifiable in edge-on systems and other mechanisms
can give rise to axisymmetric B/PS bulges (e.g.\ Binney \& Petrou
1985; Rowley 1988), the particular gaseous kinematics of B/PS bulges
further supports the view that they are simply thick bars viewed
edge-on (e.g.\ Kuijken \& Merrifield 1995; Merrifield \& Kuijken 1999;
Athanassoula \& Bureau 1999; Bureau \& Freeman 1999). Considering
that, if proven true, this link would argue that at least $50\%$ of
all bulges formed through bar-driven processes, it is crucial to
extend those tests to earlier type galaxies, where one might naively
think that merging is more important.

In this paper, we thus present generic N-body simulations of
bar-unstable disks developing a B/PS bulge (\S~\ref{sec:nbody}), and
positively compare them to the stellar kinematics (\S~\ref{sec:skin})
and $K$-band morphology (\S~\ref{sec:k-band}) of a sample of $30$
spiral galaxies, most with a B/PS bulge. We also present {\tt SAURON}
integral-field observations of a few B/PS bulges which show not only
the kinematic and structural signatures of bars, but also rather
homogeneous stellar populations, as expected (\S~\ref{sec:sauron}). We
conclude by supporting the importance of bar-driven evolution
(\S~\ref{sec:conclusions}).
%
%
\section[N-Body Simulations of Bars]
{N-Body Simulations of Bars\label{sec:nbody}}
\begin{figure}[t]
\centerline{\includegraphics[width=\textwidth]{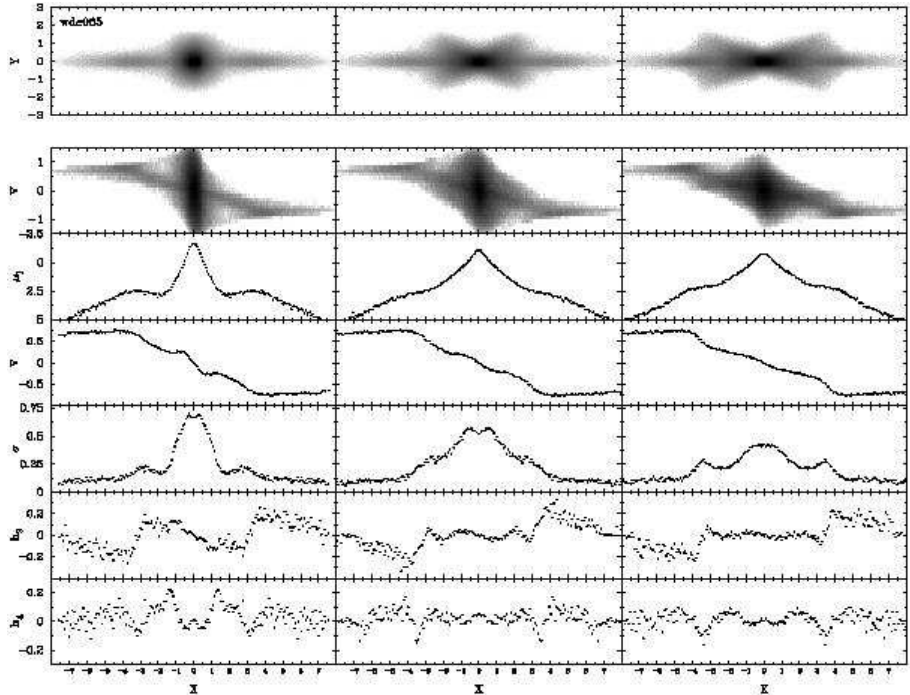}}
\caption{Bar diagnostics as a function of viewing angle for a strong
bar. From left to right: simulation seen end-on, at intermediate
viewing angle, and side-on. From top to bottom, each panel shows an
edge-on view of the simulation, the position-velocity diagram along
the major-axis, the major-axis surface brightness profile ($\mu_I$),
and the derived Gauss-Hermite coefficients $V$ (mean velocity),
$\sigma$ (velocity dispersion), $h_3$ (skewness), and $h_4$
(kurtosis). All grayscales are plotted on a logarithmic scale. Adapted
from Bureau \& Athanassoula (2004) with permission.\label{fig:nbody}}
\end{figure}

We have run a large number of standard N-body simulations of
bar-unstable disks and discuss below a few generic
results. Figure~\ref{fig:nbody} contrasts our results for a typical
strongly barred case viewed edge-on (at late times) from various
viewing angles. The initial conditions consist of a cold luminous
exponential disk with constant $Q$ and a live spherical dark halo (see
Athanassoula 2003 for more details). No luminous spheroidal component
was initially included in the simulations. The prominent thick central
component, which would normally be identified with a bulge both
morphologically and from the major-axis surface brightness profiles,
is thus composed entirely of disk material. It has acquired a large
vertical extent through disk vertical instabilities and, as expected,
appears round when seen end-on, boxy-shaped at intermediate viewing
angles, and peanut-shaped when seen side-on (e.g.\ Combes et al.\
1990).

Based on a large number of similar simulations and Gauss-Hermite fits
(van der Marel \& Franx 1993), edge-on barred disks typically show the
following kinematic signatures along their major-axis (Bureau \&
Athanassoula 2004): 1) a major-axis light profile with a
quasi-exponential central peak and a plateau at moderate radii
(Freeman Type~II profile); 2) a ``double-hump'' rotation curve; 3) a
rather flat central velocity dispersion peak with a plateau at
moderate radii and, occasionally, a local central minimum and
secondary peak; 4) an $h_3-V$ correlation over the projected bar
length. $h_4$ is rather featureless and is in any case hard to measure
observationally, so we do not discuss it further. Those kinematic
features are all spatially correlated and can be understood from the
orbital structure of barred disks. They therefore provide a reliable
and practical tool to identify bars in edge-on disks. Contrary to
popular belief, so-called ``figure-of-eight'' position-velocity
diagrams (Kuijken \& Merrifield 1995; Merrifield \& Kuijken 1999) do
not occur in the stellar kinematics, as expected for realistic orbital
configurations. However, while they are not uniquely related to
triaxiality, line-of-sight velocity distributions with a high velocity
tail (i.e.\ an $h_3-V$ correlation) do appear to be particularly
useful tracers of bars. All the characteristic kinematic features
identified grow in strength as the bar evolves and do not change
significantly for small departures from edge-on. Most can provide
useful measurements of the bar length.
%
%
%
%
\section[Stellar Kinematics of Boxy Bulges]
{Stellar Kinematics of Boxy Bulges\label{sec:skin}}
As shown in Chung \& Bureau (2004), we have obtained similar long-slit
stellar kinematics along the major-axis of the $30$ edge-on spirals
from the sample of Bureau \& Freeman (1999), most of which have a B/PS
bulge. Comparing those profiles with the N-body bar diagnostics
(\S~\ref{sec:nbody}), we find bar signatures in $80\%$ of the
galaxies, including the S0s. The diagnostics thus appear robust and
support the formation of most B/PS bulges through bar thickening.
%
%
\begin{figure}[ht]
\centerline{\includegraphics[width=0.82\textwidth]{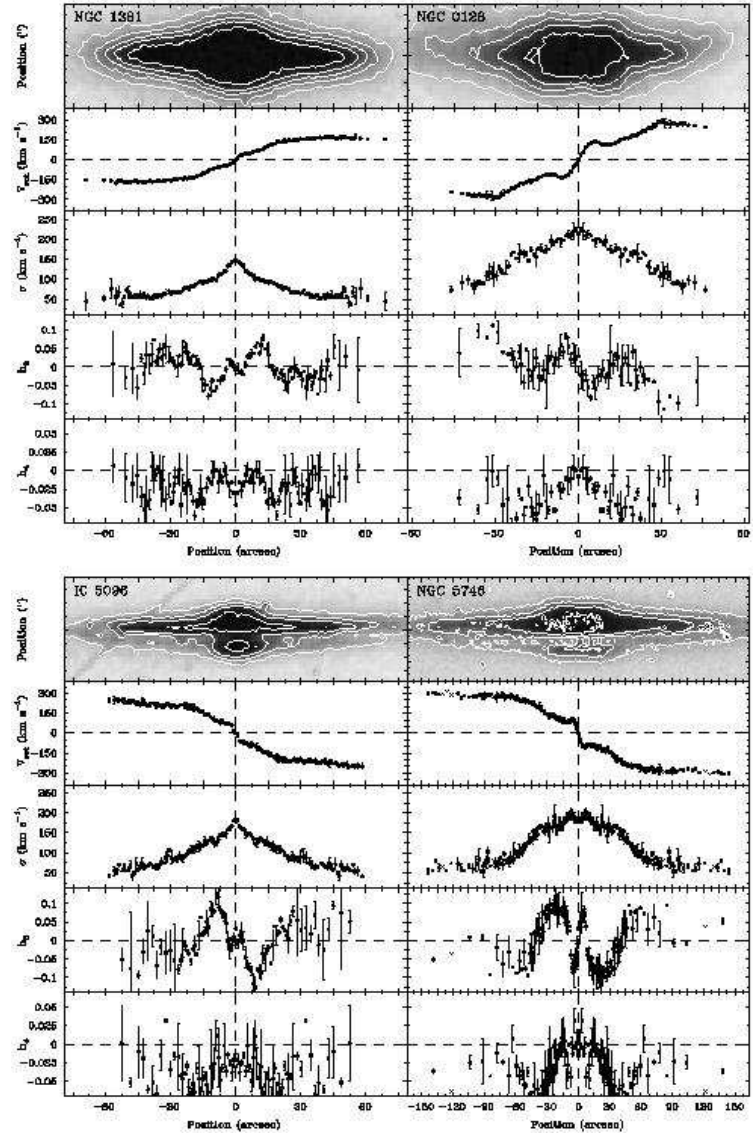}}
\caption{Stellar kinematics of spiral galaxies with a boxy bulge. The
galaxies shown are the gas-poor S0 galaxies NGC1381 (top-left) and
NGC128 (top-right) and the gas-rich intermediate-type spirals IC5096
(bottom-left) and NGC5746 (bottom-right). From top to bottom, each
panel shows an optical image of the galaxy from the Digitized Sky
Survey and the registered $V$, $\sigma$, $h_3$, and $h_4$ profiles
along the major-axis. The measurements were folded about the center
and the errors represent half the difference between the approaching
and receding sides. Adapted from Chung \& Bureau (2004) with
permission.\label{fig:skin}}
\end{figure}

As predicted, galaxies with a B/PS bulge frequently show a double-hump
rotation curve with an intermediate dip or plateau, they often display
a rather flat central velocity dispersion profile with a secondary
peak or plateau, and a significant fraction of the objects have a
local central $\sigma$ minimum ($\ge40\%$). The $h_3$ profiles display
up to three slope reversals and, most importantly, $h_3$ is normally
correlated with $V$ over the presumed bar length, contrary to
expectations from an axisymmetric disk. Figure~\ref{fig:skin} shows
the derived stellar kinematics for $4$ objects. The characteristic bar
signatures strengthen the case for an intimate relationship between
B/PS bulges and bars, even for early-type systems, and they leave
little room for competing explanations of the bulges' shape (e.g.\
Binney \& Petrou 1985). We also find that $h_3$ is anti-correlated
with $V$ in the very center of most galaxies ($\ge60\%$), suggesting
that those objects additionally harbor cold and dense (bright)
(quasi-)axisymmetric central stellar disks.

Those disks may be related to the steep central light profiles
observed (\S~\ref{sec:k-band}), and they roughly coincide with
previously identified star-forming ionized-gas disks (Bureau \&
Freeman 1999). They thus may well have formed out of gas accumulated
by the bar at its center through inflow. As suggested by N-body
models, the skewness of the velocity profile ($h_3$) appears to be a
reliable tracer of asymmetries, allowing to discriminate between
axisymmetric and barred disks seen in projection. Based on their
kinematics, B/PS bulges (and thus a large fraction of all bulges)
appear to be made-up mostly of disk material which has acquired a
large vertical extent through bar-driven instabilities, although we
have not yet probed the potentials of the galaxies out of the disk
plane systematically (but see \S~\ref{sec:sauron}). Our observations
are nevertheless consistent with standard bar-driven evolution models,
and the formation of B/PS bulges does appear to be dominated by
secular evolution processes rather than merging.
%
%
\section[$K$-Band Imaging of Boxy Bulges]
{$K$-Band Imaging of Boxy Bulges\label{sec:k-band}}
\begin{figure}[t]
\centerline{\includegraphics[width=0.96\textwidth]{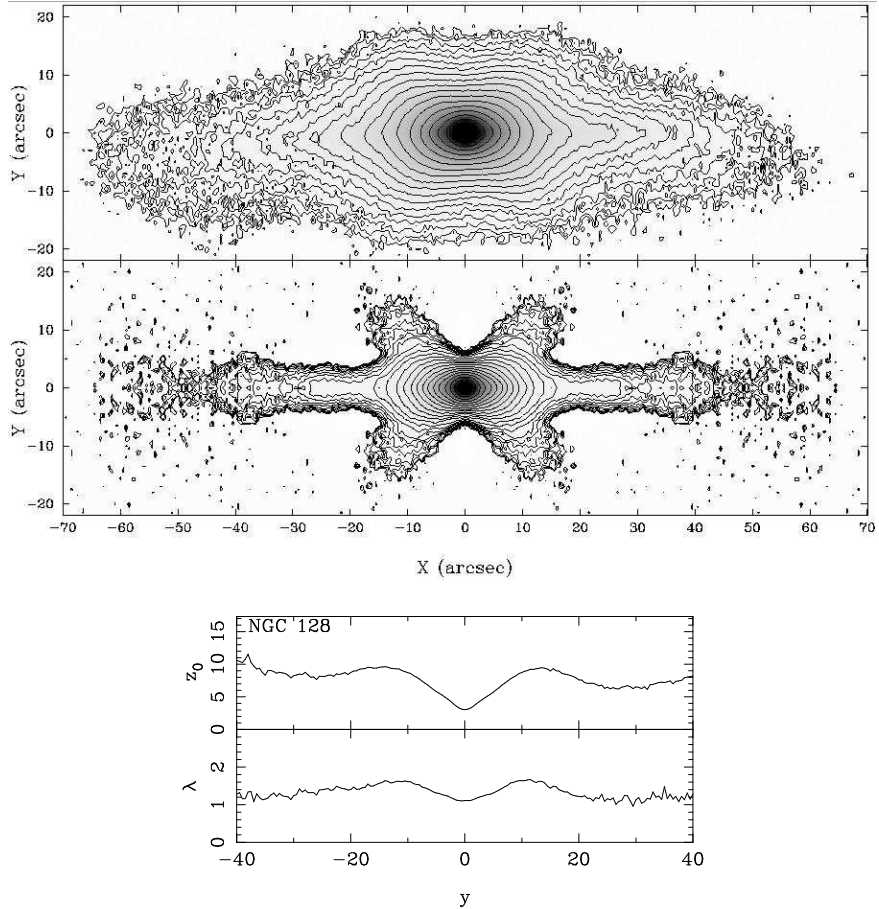}}
\centerline{\includegraphics[width=0.325\textwidth,angle=270.0]{bureau_fig3b.ps}}
\caption{Morphology and structure of spiral galaxies with a boxy
bulge. Top: $K$-band image of the S0 galaxy NGC128, showing the strong
peanut-shape of the bulge. The isophotes are separated by
$0.5$~mag~arcsec$^{-2}$. Middle: Symmetrized unsharp-masked
(median-filtered) image of NGC128, revealing the underlying orbital
structure. Bottom: Scaleheight ($z_0$) and shape ($\lambda$) of the
fitted vertical profiles of NGC128 as a function of projected radius
(registered). The radially varying thickening of the disk/bulge
material is clearly visible. Adapted from Aronica et al.\ (2005) and
Athanassoula et al.\ (2005) with permission.\label{fig:k-band}}
\end{figure}

We have also obtained $K$-band images for all the galaxies in our
sample. The $K$-band observations penetrate the prominent dust lanes
present in many galaxies and offer a much improved view of their
structure and morphology, apparently directly constraining the orbital
structure of the objects. Indeed, as illustrated in
Figure~\ref{fig:k-band} for NGC128, unsharp-masking of those images
reveals features entirely analogous to those expected from the orbital
families thought to dominate 3D bars (e.g.\ Patsis, Skokos, \&
Athanassoula 2002). In particular, the $x_1$ family ``tree'' is
clearly seen and many galaxies show secondary enhancements along the
major-axis (see Skokos, Patsis, \& Athanassoula 2002).

Moreover, the key aspect of bar thickening mechanisms to form B/PS
bulges is that the disk material is rearranged vertically (as well as
radially) by instabilities, rather than new material being added (as
expected for accretion scenarios). This process is strongly supported
by our observations, as most galaxies show a clear increase of the
scaleheight where the B/PS bulge reaches its maximum extent. This is
illustrated again in Figure~\ref{fig:k-band} for NGC128, where we
fitted the vertical profiles at each (projected) radial position with
a generalized Gaussian (equivalent to a Sersic law with
$n=\lambda^{-1}$), following Athanassoula \& Misiriotis (2002). The
shape of the profiles also changes with radius, the profiles being
shallower where the peanut shape is maximum. We note that the profiles
never approach a deVaucouleurs law ($n=4$), even in the center,
arguing again against violent relaxation and merging. Furthermore, in
many galaxies, the steep part of the light profile is much shorter
than the vertically-extended component, so that those two definitions
of a bulge are clearly inconsistent (as is that of a kinematically hot
sub-system, since B/PS bulges are most likely rotationally
supported). The steep part of the profiles is thus probably more
closely related to the central disks discussed in the previous
section.
%
%
%
%
\section[{\tt SAURON} Observations of Boxy Bulges]
{{\tt SAURON} Observations of Boxy Bulges\label{sec:sauron}}
The {\tt SAURON} team has conducted a survey of the 2D stellar
kinematics, ionized-gas kinematics, and absorption linestrengths of a
representative sample of nearby early-type galaxies (e.g.\ de Zeeuw et
al.\ 2002). The {\tt SAURON} data on edge-on early-type spirals thus
offer a unique opportunity to extend the kinematic tests described
above (\S~\ref{sec:nbody}) out of the disk plane and to test the
predictions of bar-driven evolution scenarios regarding stellar
populations.

\begin{figure}[t]
\centerline{\includegraphics[width=\textwidth]{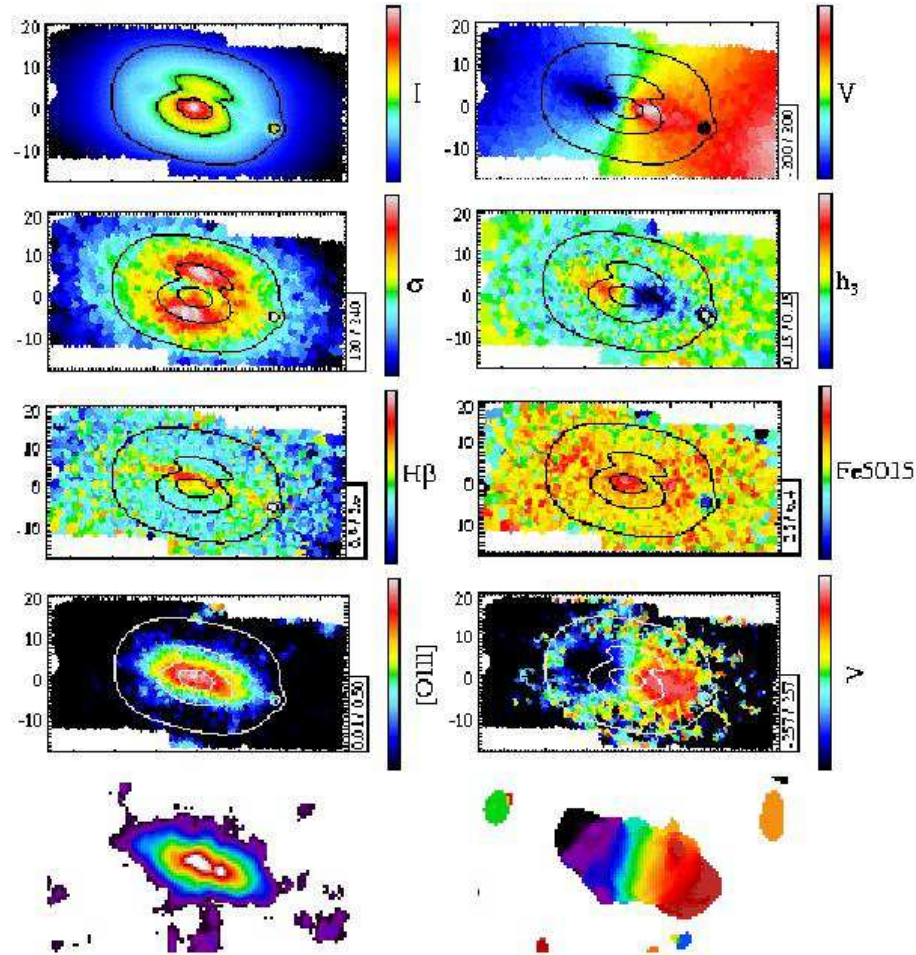}}
\caption{{\tt SAURON} and BIMA observations of the S0 galaxy
NGC4526. From left to right, top to bottom: Reconstructed image, mean
stellar velocity $V$, stellar velocity dispersion $\sigma$, asymmetry
of the stellar velocity profile $h_3$, H$\beta$ linestrength (age),
Fe5015 linestrength (metallicity), [OIII] intensity, ionized-gas
velocity, total CO flux, and CO velocity field.\label{fig:sauron}}
\end{figure}

The best examples are NGC7332 (Falc$\acute{\textrm{o}}$n-Barroso et
al.\ 2004) and NGC4526 (Fig.~\ref{fig:sauron}; Emsellem et al.\ 2004;
Sarzi et al.\ 2005; Kuntschner et al.\ 2005). Both S0 galaxies have a
boxy bulge and clearly show the stellar kinematic signatures of a
bar. They also possess homogeneous stellar populations (age and
metallicity) across the disk and bar/bulge components, except in the
central parts, as expected from simple bar-driven evolution models
(e.g.\ Friedli \& Benz 1995). NGC4526 also clearly shows a cold
central stellar disk embedded in its triaxial bulge, as traced by a
strong pinching of the stellar isovelocities in the center, a wide
local central $\sigma$ minimum, and a strong central $h_3-V$
anti-correlation (while the rest of the bulge displays an $h_3-V$
correlation as expected from a thick bar). This central stellar disk
coexists with an ionized and molecular gas disk (Young et al.\ 2005),
presumably formed through bar-driven inflow (and ensuing star
formation). NGC4526 thus beautifully illustrates and confirms most
aspect of bar-driven secular evolution models in spiral galaxies.
%
%
%
%
\section[Conclusions]
{Conclusions\label{sec:conclusions}}
As secular processes for the evolution of galaxies gain in respect and
popularity, developing practical tools to test the various models and
gain novel insights into the structure and dynamics of real galaxies
becomes increasingly pressing. The stellar kinematic bar diagnostics
presented in \S~\ref{sec:nbody} allow to identify edge-on bars easily
and thus to test the origin of B/PS bulges through bar-driven vertical
instabilities. The spectroscopic observations of a large sample of
spiral galaxies with a B/PS bulge shown in \S~\ref{sec:skin} vindicate
the usefulness of those diagnostics and confirm that B/PS bulges are
generally consistent with the presence of a thick bar viewed edge-on,
even for the earliest types. They also suggest the presence of rapidly
rotating central stellar disks at the center of the bars, which may
well have formed through bar-driven inflow and subsequent star
formation. Analysis of $K$-band observations of the same sample
reveals the orbital backbone of B/PS bulges (\S~\ref{sec:k-band}), in
agreement with expectations from periodic orbit calculations of 3D
bars. As expected, the scaleheight (and shape) of the material varies
rapidly with (projected) radius, and it is largest where the extent of
the B/PS bulge is maximum.

The emerging bar-driven evolutionary scenario is beautifully confirmed
by synthesis CO observations and {\tt SAURON} integral-field
spectroscopy (\S~\ref{sec:sauron}), which further allow to study the
distribution of the stellar populations (luminosity-weighted age and
metallicity). As the quality of stellar population data is rapidly
improving, a parallel improvement of the model predictions (which
remain rudimentary) is urgently needed. Interestingly, although clear
observational diagnostics of such scenarios are still largely
inexistent, our observations do not seem to require nor indicate that
bars may be destroyed (and possibly reformed). Specific tests would
thus also be appreciated.
%
%
\begin{acknowledgments}
M.B.\ acknowledges support by NASA through Hubble Fellowship grant
HST-HF-01136.01 awarded by the Space Telescope Science Institute. The
authors wish to thank J.C.\ Lambert, K.C.\ Freeman, and E.\ Emsellem
for support and useful discussions, and L.\ Young and the {\tt SAURON}
Team for data prior to publication. The Digitized Sky Survey was
produced at the Space Telescope Science Institute under
U.S. Government grant NAG W-2166.

\end{acknowledgments}
%
%

%
\end{document}